\documentclass[non-acm]{acmart}

\usepackage[english]{babel}
\usepackage{soul}
\usepackage{multirow}
\usepackage{booktabs} 
\usepackage{comment}
\usepackage{colortbl} 
\usepackage{array}    

\usepackage{colortbl}
\usepackage{array}
\usepackage{fontawesome5} 

\definecolor{TakeawayColor}{RGB}{60, 60, 60}   
\definecolor{TakeawayBack}{RGB}{245, 245, 245} 

\newcommand{\takeawaybox}[2]{%
    \par 
    \noindent 
    \rule{0pt}{14pt}%
    \newline 
    \begingroup
    \setlength{\arrayrulewidth}{3pt}
    \arrayrulecolor{TakeawayColor}
    \begin{tabular}{!{\vrule} >{\columncolor{TakeawayBack}} p{\dimexpr\columnwidth-15pt} @{}}
        \rule{0pt}{10pt}%
        \hspace{4pt}\textbf{\textcolor{TakeawayColor}{#1}} \newline 
        \hspace{4pt}\small #2 
        \rule[-6pt]{0pt}{12pt}%
    \end{tabular}%
    \endgroup
    \par 
    \rule{0pt}{10pt}%
    \par 
}

\AtBeginDocument{%
  }

\copyrightyear{2026}
\acmYear{2026}
\setcopyright{cc}
\setcctype{by}
\acmConference[CHASE'26]{International Conference on Cooperative and Human Aspects of Software Engineering }{April 12--18, 2026}{Rio de Janeiro, Brazil}
\acmBooktitle{International Conference on Cooperative and Human Aspects of Software Engineering (CHASE'26), April 12--18, 2026, Rio de Janeiro, Brazil}
\acmDOI{10.1145/3794860.3794899}
\acmISBN{979-8-4007-2484-8/2026/04}

\begin{document}

\title[]{Constructive Patterns for Human-Centered Tech Hiring}

\author{Allysson Allex Araújo}
 \affiliation{%
   \institution{Federal University of Cariri}
   \city{Juazeiro do Norte}
   \state{Ceará}
   \country{Brazil}
 }
 \email{allysson.araujo@ufca.edu.br}

 \author{Gabriel Vasconcelos}
 \affiliation{%
   \institution{Federal University of Cariri}
   \city{Juazeiro do Norte}
   \state{Ceará}
   \country{Brazil}
 }
 \email{gabriel.vasconcelos@aluno.ufca.edu.br}

\author{Marvin Wyrich}
 \affiliation{%
   \institution{Saarland University}
   \city{Saarbrücken}
   \country{Germany}
 }
 \email{wyrich@cs.uni-saarland.de}

\author{Maria Teresa Baldassarre}
 \affiliation{%
   \institution{University of Bari}
   \city{Bari}
   \country{Italy}
 }
 \email{mariateresa.baldassarre@uniba.it}

\author{Paloma Guenes}
 \affiliation{%
   \institution{Pontifical Catholic University of Rio de Janeiro}
   \city{Rio de Janeiro}
   \country{Brazil}
 }
 \affiliation{%
  \institution{University of Bari}
  \city{Bari}
  \country{Italy}
}
 \email{pguenes@inf.puc-rio.br}

\author{Marcos Kalinowski}
 \affiliation{%
   \institution{Pontifical Catholic University of Rio de Janeiro}
   \city{Rio de Janeiro}
   \country{Brazil}
 }
 \email{kalinowski@inf.puc-rio.br}


\renewcommand{\shortauthors}{Araújo et al.}


\begin{abstract}
[Context] Online Recruitment and Selection (R\&S) processes are often the first point of contact between early-career software engineers and the tech industry. Yet many candidates experience these processes as opaque, inefficient, or even discouraging.
While prior research has extensively documented the flaws and biases in online tech hiring, little is known about the practices that create positive candidate experiences. [Objective \& Method] This paper explores such practices, referred to as \emph{Constructive Patterns} (CPs), from the perspective of early-career software engineers. Guided by Applicant Attribution-Reaction Theory, we conducted 22 semi-structured interviews in which participants collectively described over 470 online R\&S experiences. [Results] Through thematic analysis, we identified 22 CPs that reflect positive practices such as comprehensive and transparent job advertisements (CP01), specific and developmental feedback (CP03), humanized and respectful interaction (CP06), and framing the process as a two-way street (CP18). [Conclusion] Our findings extend the conversation on tech hiring beyond diagnosing dysfunctions toward designing for human-centered and growth-oriented candidate experiences. The resulting catalog of CPs provides a concrete and empirically grounded resource for organizations seeking to attract and support early-career software engineers more effectively.
\end{abstract}






\maketitle

\section{Introduction}
\label{sec:introduction}

For many early-career software engineers, the journey into the profession begins with a fragmented and challenging experience. Recruitment and Selection (R\&S) processes, the primary gateway into the industry, are often characterized by anxiety and high expectations \citep{BEHROOZI2019, MCCARTHY2017}. Prior studies have already documented recurring pitfalls in technical hiring, from stressful and irrelevant assessments to the pervasive phenomenon of recruiter “ghosting” \cite{MONTANDON2021,behroozi2018can,EHLERS2015,biehl2025prestige, weisshaar2024hiring}. These dysfunctions have serious consequences, since they may undermine candidate experience, filter out talent unfairly, and weaken organizational pipelines at a time when skilled software professionals are in high demand. One promising way to frame these issues is through the notion of Anti-patterns (APs), introduced into the hiring context of early-career software engineers by \citet{SETUBAL2024}. In summary, APs refer to R\&S practices that appear functional in intent but systematically lead to negative outcomes for candidates. Each AP is an abstraction that synthesizes multiple observed practices reported by participants.

While this diagnostic lens is important, it remains incomplete. Knowing what to avoid does not reveal what organizations should build instead. Hence, one can ask: how can hiring processes be deliberately designed to foster  meaningful candidate experience? We argue that advancing this discussion requires a constructive lens for what we call \textit{human-centered tech hiring}: R\&S processes conceived not only as mechanisms for efficient filtering, but as humanized journeys that  support positive candidate experience while enabling organizations to make effective hiring decisions.



Moreover, to move from diagnosis to design, we turn to the Applicant Attribution-Reaction Theory (AART) \cite{PLOYHART2004}. AART emphasizes that candidates are not passive recipients of procedures; they actively interpret hiring events and assign causal explanations~\cite{PLOYHART2004, MCCARTHY2017}. For example, receiving clear and timely feedback may be attributed to organizational fairness and respect, fostering motivation and engagement. On the other hand, silence after an assessment may be interpreted as neglect or bias, eroding trust and discouraging participation. Therefore, we may assume that this attributional lens help us understand which practices systematically generate negative or positive candidate experiences \cite{carpenter2013improving, miles2018candidate}.


Building on this motivation, we introduce the concept of \textit{Constructive Patterns} (CPs): recurring and empirically grounded practices in R\&S that foster positive candidate attributions. We explicitly acknowledge that many CPs align with well-established principles in Human Resources and organizational psychology, including organizational justice (e.g., fair assessments and compensation), signaling theory (e.g., transparency in job advertisements and communication as cultural signals), and prior candidate experience studies. Accordingly, CPs do not claim novelty by introducing new psychological mechanisms. Rather, their contribution lies in the domain-specific synthesis and operationalization of these principles within tech hiring. By grounding CPs in empirical accounts and interpreting them through AART, we translate established theoretical constructs into actionable design elements tailored to the socio-technical realities of online R\&S in software engineering. In this sense, CPs serve as a candidate-centered approach that connects theory, empirical evidence, and practical decision-making.

Guided by this perspective, our study addresses the following \textbf{research question}: \textit{What Constructive Patterns shape positive experiences for novice software engineers in online Recruitment and Selection processes?} To answer this question, we conducted qualitative interviews with 22 early-career developers in Brazil, who collectively reported experiences from over 470 online R\&S processes. Using thematic analysis, we derived a preliminary catalog\footnote{https://gesid.github.io/papers/swe-cps-hiring} of 22 CPs that organizations can adopt as building blocks for more human-centered hiring processes. In this regard, our contributions are threefold:


\begin{itemize}
\item We move from Anti-patterns (APs) to Constructive Patterns (CPs), bridging two complementary perspectives and enabling a practical shift from identifying dysfunctions to highlighting positive practices.

\item We provide an actionable catalog of 22 CPs, presented as a practical resource that organizations may use to self-assess their hiring processes, guide redesign efforts, and establish mechanisms for continuous improvement toward human-centered tech hiring.

\item We extend the AART into the domain of software engineering hiring, offering new empirical evidence of how candidate attributions shape perceptions of fairness, engagement, and behavioral intentions.
\end{itemize}


This paper is structured as follows: Section \ref{sec:background} provides the background and Section \ref{subsec:related_work} covers the related work. Section \ref{sec:method} details the methodology. Section \ref{sec:results} presents the results and analysis. Section \ref{sec:discussion} discusses the implications and key takeaways. Section \ref{sec:threats} addresses threats to validity, and Section \ref{sec:conclusion} offers concluding remarks.

\section{Background}
\label{sec:background}

Our research builds on two core conceptual perspectives: the R\&S process as an interpretive journey for the candidate, and AART as the main theoretical lens to understand that interpretation.

\subsection{The Recruitment and Selection Process}
\label{subsec:rs_process}

The Recruitment and Selection (R\&S) process is a multi-phase journey that represents the main gateway into an organization \cite{ANDERSON2002}. We conceptualize the R\&S process as starting with candidates’ first exposure to job-related signals (e.g., job postings and recruitment channels) and extending through the offer decision. However, beyond its procedural structure, R\&S is inherently a communicative and interpretive process. Each phase sends signals that shape how candidates perceive the organization. Viewed through this lens, R\&S is more than an administrative function, it is a key social system through which organizations convey their identity and values. Furthermore, its design directly affects not only who is hired but also how future employees perceive the organization. Broadly, \textbf{recruitment} comprises the \textit{attraction} and \textit{screening} stages, while \textbf{selection} covers the \textit{assessment} and final \textit{hiring} decision. These phases form an integrated process culminating in the incorporation of new employees into the organization \cite{KAMRAN2015,SWAMY2021}.

In the \textit{attraction} phase, organizations define job requirements, profiles, and disseminate opportunities through appropriate channels, increasingly relying on online platforms and social networks \cite{HOLM2018}. Applications (e.g., résumés, cover letters) formalize the candidate’s entry into the process. The subsequent \textit{screening} phase filters applicants who meet minimum requirements, shaping the pool for later evaluation \cite{MCCARTHY2017}.  The \textit{assessment} phase involves both behavioral and technical evaluations. Behavioral assessments may include structured or semi-structured interviews, group dynamics, or psychometric tests, while technical assessments typically involve coding tasks, system design challenges, or problem-solving exercises. Although intended to evaluate competencies and cultural fit, the perceived fairness and relevance of these assessments strongly shape candidate experience \cite{ANDERSON2002,BEHROOZI2019,PLOYHART2004}.  Finally, the \textit{hiring} phase formalizes the relationship through an offer and negotiation, where clarity of terms and overall impressions of the process play a decisive role \cite{MCCARTHY2017}. In this sense, R\&S functions as a series of sensemaking events, in which candidates actively interpret cues to form attributions about the organization.

\subsection{Applicant Attribution-Reaction Theory}
\label{subsec:aart}

Rooted in social psychology, Attribution-Reaction Theory (AART) posits that candidates are not passive recipients of procedures; they are active agents who constantly seek to understand the causes behind R\&S events \cite{PLOYHART2004}. Hence, AART positions itself as a theory that explores how candidates perceive and respond to the R\&S process based on their experience. It begins with an \textit{objective R\&S event}, which is any concrete action or occurrence in the process (e.g., receiving personalized feedback). This event triggers a \textit{causal attribution}, where the candidate assigns a reason for the event (e.g., ‘this company values my time and potential’). This attribution, in turn, shapes their \textit{perceptions} of the process and the organization (e.g., a feeling of fairness and respect). Finally, these perceptions influence subsequent \textit{behavioral reactions}, such as increased motivation or a likelihood of accepting an offer \cite{MCCARTHY2017}.
While AART explains how flawed processes lead to negative outcomes, its main value here is generative: by focusing on desired attributions (such as fairness, transparency, respect, and partnership) it shifts the goal from avoiding problems to intentionally creating positive experiences.


\section{Related Work}
\label{subsec:related_work}

Research in SE has increasingly focused on the dysfunctions within hiring processes, revealing structural and experiential challenges that affect both recruiters and candidates. \citet{BEHROOZI2019}, for instance, have provided strong evidence that technical interviews are often stressful, disconnected from real-world tasks, and poor predictors of performance. Their analysis of developer testimonials also highlighted widespread frustration with communication breakdowns, opaque expectations, and a lack of transparency throughout the hiring pipeline \cite{BEHROOZI2020SHIROLKARBARIKPARNIN_DEBUGGING}. These findings paint a picture of hiring as an experience often marked by friction and ambiguity, and emphasize the need to better understand what practices actually create effective and fair experiences for candidates.

Other studies have examined specific mechanisms at the start of the hiring funnel. \citet{MONTANDON2021} identified frequent mismatches between the skills listed in job advertisements and the actual demands of the role, creating misaligned expectations for candidates. In turn, Ehlers \cite{EHLERS2015} showed how subtle cues in job descriptions can influence applicant attraction, while \citet{fritzsch2023resist} characterized the phenomenon of Résumé-Driven Development, where both employers and candidates prioritize trending technologies over substantive job alignment. More recently, \citet{wyrich2026} examined the evolving profile of software engineers by analyzing 100 SE job postings, showing that employers increasingly emphasize non-technical attributes such as cultural fit, sense of purpose, interpersonal skills, and continuous growth.

Another stream of research has explored the hiring process from the organizational or recruiter's perspective. \citet{SETUBAL2024}, for example, introduced the concept of Anti-patterns (APs) to the hiring context, developing a catalog of recurring problematic practices as seen by experienced recruiters. Similarly, \citet{biehl2025prestige} investigated how recruiters rely on prestige-based heuristics to filter candidates, often perpetuating social inequities. \citet{brown2025towards} further advanced the discussion by outlining systemic shortcomings in tech hiring processes, including the overreliance on subjective assessments, lack of validated instruments, and inconsistent interview structures for the candidates.


The existing literature provides a strong diagnostic account of hiring flaws from various angles. However, a theoretically-grounded and constructive model of what works from the candidate's perspective is still in an initial stage. Our study addresses this gap by shifting the focus from APs to CPs. Unlike prior related work, we intentionally center the candidate's subjective experience and strategically use AART to understand \textit{why} certain practices are considered effective. 

\section{Method}
\label{sec:method}

To explore the CPs that foster positive experiences for novice software engineers in online R\&S, we adopted an exploratory qualitative methodological research approach, described in detail below. 


\subsection{Research Design}
\label{subsec:design_goal}

This research is guided by the following research question: \textit{What constructive patterns shape positive experiences for novice software engineers in online recruitment and selection processes?} Our approach is grounded in a constructivist epistemology, which posits that knowledge is co-constructed through the interactions between researchers and participants \cite{rockmore2005constructivist}. We used the AART \cite{PLOYHART2004} as our guiding theoretical lens for interpreting candidate experiences. In other words, AART helps explain \textit{why} certain recruitment practices are perceived as positive or negative by focusing on how applicants generate causal attributions about specific R\&S events. Consequently, these attributions influence their perceptions of fairness, respect, and organizational credibility, which in turn shape subsequent attitudes and behaviors. Inspired by this broader perspective, our primary objective was to identify a comprehensive catalog of CPs that may foster positive attributions, strengthening candidates’ sense of fairness, respect, and competence throughout the hiring process.

\subsection{Data Collection}
\label{subsec:instrument}

We collected primary data through semi-structured interviews, guided by a script whose full content is available in our open repository \cite{zenodo}. The script was developed based on the constructs of AART and designed to capture participants’ perceptions while allowing sufficient flexibility to explore individual narratives in depth. The core of the script explored the typical phases of an R\&S process (Application, Assessment, and Hiring). Moreover, open-ended guiding questions were formulated to target specific AART constructs, allowing us to elicit detailed narratives about the following domains:

\begin{itemize}
    \item \textbf{Objective R\&S Events:} We investigated candidates' concrete experiences with job ads, communication, feedback, and technical or behavioral assessments (e.g., “\textit{How do you evaluate the relevance of the practical activities or tests applied during the selection process?}”);
    
    \item \textbf{Causal Attributions:} A key focus was understanding how candidates explained and made sense of their positive experiences, forming the basis for identifying CPs. The script was designed to elicit them through evaluative questions (e.g., “\textit{How do you evaluate the feedback you received?”}). This approach prompted participants to openly articulate the reasons behind their perceptions of fairness, respect, or success;
    
    \item \textbf{Formed Perceptions:} We explored subjective evaluations of the process, such as perceptions of fairness, transparency, and cultural fit (e.g., “\textit{How do you perceive the fairness of the selection process used by companies?}”);
    
    \item \textbf{Manifested Reactions:} We investigated the positive impacts of these experiences, exploring how they influenced a candidate's decision to accept an offer or their positive feelings toward an organization (e.g., “\textit{What crucial factors do you consider in your decision-making process to accept or decline an offer?}”).
\end{itemize}

The semi-structured design of the interview protocol allowed the interviewer to adapt the sequence of questions, formulate follow-up prompts based on participants’ narratives, and explore underlying causes and meanings in greater depth. Before each interview, the researcher restated the study’s objectives, emphasized confidentiality, and obtained informed consent to foster trust and encourage open, detailed sharing. The interview protocol was piloted to assess clarity and procedure; because the pilot interview (P00) produced relevant data and required no substantive revisions, it was included in the final analytic corpus.



\subsection{Participant Selection and Characterization}
\label{subsec:participants_collection}

Participants were recruited via professional networks and selected based on two criteria: (1) possessing up to three years of professional experience, and (2) having participated in at least three distinct R\&S processes. A total of 22 individual interviews were conducted, each lasting between 30 and 45 minutes. All sessions were audio-recorded and later transcribed verbatim, resulting in 264 pages of text (Arial, font size 12). Table \ref{tab:participants_characterization} summarizes the participants’ profiles. They represented a range of roles (e.g., Developer, QA, DevOps) and had collectively taken part in more than 470 online R\&S processes. We determined this sample size based on information power \cite{malterud2016sample}. Hence, data collection concluded when interviews reinforced existing codes and Constructive Patterns, with no substantively new codes emerging, indicating sufficient analytical depth.

\begin{table}[t!]
\centering
\caption{Characterization of Study Participants (N=22)}
\label{tab:participants_characterization}
\resizebox{\columnwidth}{!}{%
\begin{tabular}{lllc}
\toprule
\textbf{Part.} & \textbf{Most Recent Role} & \textbf{Prof. Exp. Time} & \textbf{\shortstack{Online R\&S\\Processes}} \\ \midrule
P00 & Quality Analyst & 3 months & 3 \\
P01 & Software Developer & 2 years & 50+ \\
P02 & Software Developer & 1 year & 4 \\
P03 & Software Developer & 3 years & 15 \\
P04 & Quality Analyst & 3 years & 5 \\
P05 & Quality Analyst & 3 years & 6 \\
P06 & Quality Analyst & 2 years \& 6 months & 5 \\
P07 & Quality Analyst & 2 years & 50+ \\
P08 & Software Developer & 3 years & 6 \\
P09 & Software Developer & 3 years & 3 \\
P10 & Software Developer & 1 year \& 1 month & 50+ \\
P11 & Software Developer & 1 year & 50+ \\
P12 & Software Developer & 2 years \& 6 months & 30 \\
P13 & Process Analyst & 2 years \& 6 months & 7 \\
P14 & UI/UX Designer & 2 years \& 6 months & 3 \\
P15 & Software Developer & 1 year \& 7 months & 50+ \\
P16 & Quality Analyst & 2 years & 4 \\
P17 & DevOps Engineer & 2 years \& 5 months & 5 \\
P18 & Quality Analyst & 3 years & 10 \\
P19 & Software Developer & 1 month & 5 \\
P20 & Software Developer & 2 years & 50+ \\
P21 & Software Developer & 3 years & 9 \\ \bottomrule
\end{tabular}%
}
\end{table}


\subsection{Data Analysis}
\label{subsec:data_analysis}

To construct our catalog of CPs, we conducted a thematic analysis of the interview transcripts, drawing inspiration from the six-phase approach to thematic analysis described by Braun and Clarke \cite{braun2006using}. Our analysis was inductive and reflexive, designed with the goal of identifying, defining, and characterizing the CPs reported by early-career software engineers. The process followed these steps:

\begin{enumerate}
    \item \textbf{Data Familiarization:} The first phase consisted of an immersion in the data. We read and reread all 22 transcripts to develop a holistic understanding of the participants' narratives. During this process, our analytical lens was sensitized to identify instances of satisfaction, praise, respect, fairness, or positive engagement, which were provisionally highlighted for subsequent coding.

    \item \textbf{Generating Initial Codes:} We then performed an inductive open coding process focused on the highlighted positive excerpts. Each distinct positive experience, organizational action, or perception was captured with a descriptive code. For example, a participant's statement about receiving clear salary information in a job advertisement was assigned the code ``clarity on core job conditions''. This bottom-up approach generated a comprehensive list of initial codes directly grounded in the participants' own words.

    \item \textbf{Searching for Themes:} The initial codes were then organized within a thematic structure. This structure was derived from the AART-informed interview script, which already segmented the R\&S journey into logical phases (e.g., Application, Assessment, etc.) and cross-cutting dimensions (e.g., Feedback, Perception of Justice, etc.). 
    

    \item \textbf{Reviewing and Refining Codes:} This iterative phase involved a systematic review to ensure conceptual robustness. Initial coding generated 39 descriptive codes, which were reviewed, refined, and collapsed into the 22 resulting CPs. We employed a constant comparison method, examining emerging data against existing codes to merge conceptual overlaps, for instance, distinct initial codes regarding salary transparency'', ``tech stack clarity'', and ``project context'' (mentioned by P09, P01 and P12) were synthesized into the broader pattern CP01 (\textit{Comprehensive and Transparent Job Advertisement}). In turn, saturation was judged based on the recurrence of these patterns across the diverse participant pool; as analysis progressed, no new initial codes emerged that could not be mapped to the existing structure.

    \item \textbf{Defining and Naming Constructive Patterns:} In this phase, the refined and validated codes were abstracted into the final 22 CPs. This step involved moving from descriptive codes to conceptual patterns. Each CP was given a descriptive name (e.g., \textit{CP01: Comprehensive and Transparent Job Advertisement}) and a definition that captured its principle and underlying mechanism. This entire process was documented in a detailed, publicly available codebook \cite{zenodo} to ensure transparency and traceability.

    \item \textbf{Producing the Report:} The final step was the construction of the analytical narrative presented in this paper. For each CP, we selected representative quotes from our codebook to serve as empirical evidence. We then interpreted the significance of each pattern through the lens of AART, explaining how the practice likely fosters the positive attributions.
\end{enumerate}

\subsection{Methodological Trustworthiness}
\label{subsec:rigor}

To ensure the trustworthiness of our findings, we adopted rigorous strategies from qualitative research \cite{nowell2017thematic, gioia2013seeking}. \textbf{Auditability} was maintained through a transparently documented analytical process and the public availability of our research artifacts \cite{zenodo}. To mitigate single-coder bias, data analysis involved four researchers with experience in qualitative research. While initial coding was conducted by one author, subsequent phases were collaboratively reviewed through \textbf{peer debriefing} sessions. When interpretive disagreements emerged during code refinement, they were resolved through iterative discussion. \textbf{Theoretical coherence} was maintained by consistently grounding our interpretation of the CPs in the principles of AART, ensuring alignment between data, theory, and emergent findings. Finally, we used \textbf{thick descriptions} by providing contextualized quotes to allow readers to assess the validity of our interpretations and enhance the transferability of the findings. These measures aimed to establish methodological rigor and trustworthiness throughout the study.



\section{Results and Analysis}
\label{sec:results}


Our thematic analysis identified recurring CPs that characterize a constructive, human-centered hiring experience for novice software engineers. We synthesized these into a catalog of 22 CPs, observable practices consistently described by candidates as prompting positive attributions. Accordingly, CPs reflect how hiring practices are experienced and interpreted by candidates, rather than how often or effectively organizations implement them. Table~\ref{tab:cp_framework} provides an overview of the catalog, while the full codebook with participant traceability is available in our repository \cite{zenodo}. In the subsections that follow, we expand on each CP, illustrating how it operates in practice and interpreting its impact through the lens of AART.


\begin{table}[t!]
\centering
\caption{Resulting Catalog of 22 Constructive Patterns (CPs) for human-centered online tech hiring
}
\label{tab:cp_framework}
\resizebox{\columnwidth}{!}{%
\begin{tabular}{ll}
\toprule
\textbf{ID} & \textbf{Constructive Pattern} \\ 
\midrule
CP01 & Comprehensive and Transparent Job Advertisement \\
CP02 & Verified Positive Reputation \\
CP03 & Specific and Developmental Feedback \\
CP04 & Relevant and Purposeful Technical Assessment \\
CP05 & Direct Interaction with the Team and Managers \\
CP06 & Humanized and Respectful Interaction \\
CP07 & Visible Commitment to Diversity, Inclusion, and Accessibility \\
CP08 & Signals of Long-Term Stability and Growth \\
CP09 & Co-creation of Job Descriptions with Technical Teams \\
CP10 & Streamlined and Efficient Application Process \\
CP11 & Communication as a Proxy for Company Culture \\
CP12 & Soliciting Candidate Feedback on the R\&S Process \\
CP13 & Timely and Efficient Process Pacing \\
CP14 & Continuous and Transparent Status Updates \\
CP15 & Proactive Candidate Redirection \\
CP16 & The Interview as a Learning Experience \\
CP17 & Fair Compensation Aligned with Requirements \\
CP18 & The Process as a Two-Way Street \\
CP19 & Proactive Talent Pooling \\
CP20 & Technically Proficient Interviewers \\
CP21 & Effective Use of Automation for Efficiency \\
CP22 & Holistic Candidate Evaluation \\ 
\bottomrule
\end{tabular}%
}
\end{table}


\subsection{The Application Phase}
The first point of contact between a candidate and an organization is the application phase. The practices employed here are critical, as they shape initial perceptions and determine whether a candidate will invest time and effort. Our findings show that CPs in this phase build a foundation of trust through transparency, credibility, and respect. Table \ref{tab:app_phase_cps} summarizes the CPs emerged in this phase.

\begin{table}[!ht]
\centering
\caption{Constructive Patterns within the Application Phase}
\label{tab:app_phase_cps}
\begin{tabular}{ll}
\toprule
\textbf{Theme} & \textbf{Associated CPs} \\ \midrule
T1. Clarity in Job Postings & CP01, CP09 \\
T2. External Information & CP02 \\
T3. Feedback & CP03, CP12, CP15, CP19 \\ \bottomrule
\end{tabular}
\end{table}

\subsubsection{\textbf{T1. Clarity in Job Postings}}
The job advertisement represents a signal of an organization's professionalism and respect \cite{EHLERS2015, MONTANDON2021}. This initial touchpoint shapes a candidate's attributions about the company's competence and honesty \cite{PLOYHART2004}. In this context, our analysis identified two CPs: ensuring the ad is comprehensive (CP01) and involving technical teams in its creation (CP09).

\paragraph{CP01: Comprehensive and Transparent Job Advertisement}
Within AART, transparency in a job ad functions as an objective event that fosters attributions of \textit{honesty} and \textit{professionalism}. This CP refers to job ads that state core conditions like salary range, provide a description of responsibilities, and offer project context, thereby respecting the candidate's time. For instance, P09 emphasized the value of financial clarity: \textit{“Man, the salary... it's a very crucial point, because then you can already have a basis”}. In a competitive market where candidates must be selective, such details are invaluable. This issue was also echoed by P12, who highlighted the need for project details to assess fit: \textit{“it would be very good if they included... information about the context in which the person will work”}.

\paragraph{CP09: Co-creation of Job Descriptions with Technical Teams}
Candidates perceive job descriptions as most accurate when crafted with input from the hiring team. Such collaboration encourages attributions of \textit{competence} and \textit{organizational alignment}. This view was shared by P16, who reinforced the need for this synergy: \textit{“I think it's essential to have someone from the department... helping to put together these aspects of the ad”}. When an ad is technically sound, candidates infer that the organization values expertise, as P17 noted: \textit{“...where the person writing the vacancy is someone from the field... Then it's usually a better-described vacancy”}.

\subsubsection{\textbf{T2. External Information}}
The R\&S journey is not limited to official channels; candidates act as proactive investigators who seek external validation to inform their decisions \cite{MCCARTHY2017}. This search for external social proof serves as a contextual factor in AART, shaping attributions of trustworthiness even before direct interaction occurs. This proactive evaluative process is captured in the importance of a company's verified positive reputation (CP02).

\paragraph{CP02: Verified Positive Reputation}
We have noticed that candidates seek external validation on platforms like Glassdoor or through informal networks. Hence, a strong reputation acts as a motivator, leading to pre-emptive attributions of the organization being a \textit{trustworthy} and \textit{desirable} employer. P05 explained this investigative mindset: \textit{“It's Glassdoor... I always do some research beforehand... sometimes I go on LinkedIn and I send a message to someone who works there”}. Such findings directly impact motivation and trust, as P20 stated: \textit{“I believe the two strongest... are Glassdoor and LinkedIn. Glassdoor has the mechanic of anonymous reviews from other employees, including salary”}.

\subsubsection{\textbf{T3. Feedback}}
Feedback emerged as another important dimension of the candidate experience, capable of transforming a potential rejection into a constructive interaction \cite{ANDERSON2002}. Within AART, the act of giving feedback could be interpreted as an ``objective event'' that triggers attributions of fairness and respect \cite{PLOYHART2004}. Our analysis surfaced CPs that build this trust, from providing specific feedback (CP03) and soliciting their input on the process (CP12), to more advanced issues like proactively redirecting strong candidates (CP15) and maintaining a talent pool (CP19).

\paragraph{CP03: Specific and Developmental Feedback}
This CP involves providing personalized and actionable feedback that explains the reasons for a decision. It fosters strong attributions of \textit{fairness}, \textit{respect}, and investment in the candidate's growth. For example, P13 noted the value of understanding the reason for a decision: \textit{“it's cool to receive personalized feedback, right? [...] The reason you were stopped at that phase”}. In a market where interview standards are rising, such guidance becomes even more important. Some organizations go further; P04 even received \textit{“study material and everything”}, transforming feedback into a direct learning opportunity.

\paragraph{CP12: Soliciting Candidate Feedback on the R\&S Process}
This CP involves proactively asking candidates for their own feedback on the hiring experience. We may notice that this act signals that the candidate's perspective is valued and that the organization is committed to improvement. Thus, it shifts the power dynamic from a one-way evaluation to a mutual dialogue, fostering attributions of \textit{partnership} and \textit{organizational reflexivity}. P10 clarified this feeling: \textit{“I'm also important and I also have a voice here to say if the process was cool... And you can feel that”.}

\paragraph{CP15: Proactive Candidate Redirection}
Another identified CP is the proactive redirection of a strong candidate to a more suitable role. This approach transforms a negative outcome into a new opportunity, generating attributions of \textit{respect} and a genuine \textit{interest in the candidate's talent}. P04, who experienced this scenario, was given \textit{“the opportunity to apply to participate in another interview... for another vacancy [...] that my profile matched better”}.

\paragraph{CP19: Proactive Talent Pooling}
This CP informs a promising candidate that their profile will be kept in a talent database for future opportunities. This decision softens the rejection by offering a sense of future possibility, signaling that the company still values the candidate's profile. This perspective leads to attributions of \textit{respect} and \textit{foresight}, as confirmed by P16: \textit{“I've already been called for an interview because of these resume databases”}.



\subsection{The Assessment Phase}
\label{subsec:assessment_phase}

The Assessment Phase is often the most high-stakes and anxiety-inducing part of the R\&S process, where candidates must demonstrate their competence under scrutiny. Our findings reveal that positive experiences in this phase are not about making assessments easier, but about designing them to be fairer, more relevant, and more human. CPs in this phase focus on validating a candidate's skills while respecting them as a professional, thereby strengthening attributions of procedural and interpersonal justice \cite{GILLILAND1993, PLOYHART2004}. Table~\ref{tab:assess_phase_cps} outlines the CPs identified in this phase.

\begin{table}[!ht]
\centering
\caption{Constructive Patterns within the Assessment Phase}
\label{tab:assess_phase_cps}
\begin{tabular}{ll}
\toprule
\textbf{Theme} & \textbf{Associated CPs} \\ \midrule
T4. Behavioral Assessment & CP22 \\
T5. Technical Assessment & CP04, CP16, CP20 \\ \bottomrule
\end{tabular}
\end{table}

\subsubsection{\textbf{T4. Behavioral Assessment}}
While technical proficiency is critical in software engineering, our participants also valued processes that recognized their human qualities. Constructive experiences in behavioral assessment were not tied to specific tools but to the underlying principle of seeing the candidate as a multi-faceted individual. This perception is captured in the commitment to a holistic evaluation (CP22).

\paragraph{CP22: Holistic Candidate Evaluation}
This CP describes an assessment process that shows genuine interest in the candidate as a person, including soft skills and personality. This human-centered approach triggers attributions of \textit{fairness}, signaling that the company values diverse contributions. P14 viewed this as a just practice: \textit{“they wanted to know the person, really soft skills, really in terms of cultural fit... So this I consider something fair, you know?”} This approach avoids reducing a candidate to technical keywords, a sentiment P18 appreciated: \textit{“they want to know... how you are as a person, to know if you will fit into the company's ecosystem”}.

\subsubsection{\textbf{T5. Technical Assessment}}
The technical evaluation is arguably the most scrutinized stage in tech hiring, often criticized for being stressful and disconnected from real-world practice \cite{BEHROOZI2019}. Our findings, however, point to a constructive alternative. Rather than an interrogation, a positive technical assessment is perceived as a relevant, collaborative, and credible exchange. This perspective is observed through three interconnected CPs: designing relevant assessments (CP04), framing the interview as a learning experience (CP16), and ensuring interviewers are technically proficient (CP20).

\paragraph{CP04: Relevant and Purposeful Technical Assessment}
This CP refers to technical assessments related to the daily tasks of the role. When a test mirrors the job, candidates perceive it as a valid measure of their suitability, fostering attributions of \textit{procedural justice}. As P15 noted, project-based tests are more engaging: \textit{“...technical tests of projects themselves... focused really on how your day-to-day work would be... I think those are interesting”}. This issue aligns with an industry trend where startups seek practical, “cracked” engineers \cite{orosz2025state}. The assessment must also be appropriate for the candidate's seniority, as P04 highlighted: \textit{“it was consistent with what the vacancy asked for, and with my knowledge too, with the seniority issue”}.

\paragraph{CP16: The Interview as a Learning Experience}
This CP reframes the interview from a one-sided test into a collaborative exchange that allows the candidate to learn something new. This CP generates attributions of a \textit{supportive} and \textit{development-oriented} culture. P04 described such an interview as \textit{“something very dynamic, it was even something that I left with more knowledge than I entered”}. P21 also highlighted the value of collaboration during the interview, noting how \textit{“you get new ideas, how you can implement things, even for your portfolio”}, thus turning the evaluation into a developmental moment. This strengthens their positive attribution towards the company's brand, making them more likely to reapply or recommend it.

\paragraph{CP20: Technically Proficient Interviewers}
The credibility of the technical assessment hinges on the perceived competence of the interviewer. With hiring bars reportedly rising \cite{orosz2025reality}, candidates feel their skills can only be fairly evaluated by a knowledgeable expert. This view fosters attributions of \textit{validity}, \textit{fairness}, \textit{trust} and \textit{respect}. As P03 stated, \textit{“the best interviews I've ever had are when we have a chat with the tech area... a tech leader from some team comes”}. The perception of expertise lends legitimacy to the process, as P19 contrasted: a good interviewer \textit{“gives me the notion that they know what they're asking for”}.



\subsection{The Hiring Phase}
\label{subsec:hiring_phase}

The final steps of the R\&S process, such as the job offer, make an organization’s promises tangible. CPs at this stage focus on distributive justice, outcome fairness, and clearly communicating the candidate’s future within the company. These CPs are key to turning a positive candidate experience into a successful hire. Table~\ref{tab:hiring_phase_cps} summarizes them.

\begin{table}[!ht]
\centering
\caption{Constructive Patterns within the Hiring Phase}
\label{tab:hiring_phase_cps}
\begin{tabular}{ll}
\toprule
\textbf{Theme} & \textbf{Associated CPs} \\ \midrule
T6. Benefits and Opportunities & CP08, CP17 \\ \bottomrule
\end{tabular}
\end{table}

\subsubsection{\textbf{T6. Benefits and Opportunities}}
At the offer stage, clarity about compensation and future prospects is paramount. A positive experience here depends on the organization's ability to present a compelling and fair value proposition, addressing both immediate rewards and long-term career potential.

\paragraph{CP08: Signals of Long-Term Stability and Growth}
This CP involves the company providing clear information that allows candidates to envision a future, not just a job. This generates attributions of the organization as a \textit{stable} and \textit{supportive} place that invests in its people. As P05 noted, a key question is whether a junior can \textit{“become a senior... I want to see the growth opportunity”}. This extends to development, as P13 valued the \textit{“possibility of growth and rotation in the company, which is having the possibility to study other areas”}.

\paragraph{CP17: Fair Compensation Aligned with Requirements}
This CP describes a compensation package perceived as fair and proportional to the required skills. In other words, a fair offer leads candidates to attribute \textit{distributive justice} and \textit{market awareness} to the organization. P06 summarized this expectation: \textit{“I think the salary issue is important, if the salary is in accordance with the job requirements”}. For candidates, this is about recognizing their investment in their own skills, as P07 explained: \textit{“I have certification, I have education... so I will want a salary compatible with that”}.



\subsection{Cross-Cutting Aspects}
\label{subsec:cross_cutting_aspects}

Beyond the distinct phases of R\&S, our analysis also revealed a set of CPs that represent cross-cutting aspects. These CPs are not confined to a single stage but span the entire R\&S journey, shaping the candidate's perception of an organization's character, competence, and values. Table~\ref{tab:cross_phase_cps} outlines these CPs.

\begin{table}[!ht]
\centering
\caption{Constructive Patterns within Cross-Cutting Aspects}
\label{tab:cross_phase_cps}
\begin{tabular}{ll}
\toprule
\textbf{Theme} & \textbf{Associated CPs} \\ \midrule
T7. Conduct of R\&S Process & CP05, CP06, CP10, CP11, \\
& CP13, CP18, CP21 \\
T8. Cultural Fit & CP01, CP03, CP05, CP06, CP07, \\
& CP11, CP22 \\
T9. Process Status & CP14 \\ 
T10. Perception of Justice & CP07 \\
\bottomrule
\end{tabular}
\end{table}

\subsubsection{\textbf{T7. Conduct of the R\&S Process}}
This theme captures the overall careful orchestration of the hiring journey. Constructive experiences are defined by processes that are humanized, efficient, and  transparent in their execution, signaling a respectful organization.

\paragraph{CP05: Direct Interaction with the Team and Managers}
This CP highlights the opportunity for candidates to have meaningful conversations with future colleagues. This stance is perceived as the most authentic way to assess fit, fostering attributions of \textit{transparency} and \textit{peer-level respect}. As P00 noted, the chance to \textit{“chat with your probable future project manager... weighs heavily on the professional's decision”}. P21 also added that it allows one to \textit{“get to know the personalities of the people there”}.

\paragraph{CP06: Humanized and Respectful Interaction}
This CP describes interactions where recruiters and interviewers treat candidates with respect and genuine interest. Such thoughtful interactions are a direct source for attributions of \textit{interpersonal justice} and \textit{organizational care}. As P10 described, \textit{“I felt a connection, and like... humanization from a company that at the same time is gigantic”}. The positive and lasting impact of such treatment was confirmed by P08, who, reflecting on all interactions from the recruiter to management, described the experience as \textit{“very satisfactory”}.

\paragraph{CP10: Streamlined and Efficient Application Process}
Candidates value application methods that are quick and respect their time. This efficiency is perceived as a sign of a \textit{modern} and \textit{candidate-centric} organization. P03 stated a clear preference: \textit{“I concentrate everything on LinkedIn, mainly on vacancies that are simplified”}. P14 shared this view, stating, \textit{“I can use it on LinkedIn... and have the simplified application, which I like a lot”}.

\paragraph{CP11: Communication as a Proxy for Company Culture}
We found that candidates often perceive the quality of communication as a reliable indicator of the company's internal culture. Then, positive and professional communication leads to attributions of a \textit{healthy} and supportive work environment. P04 explained this inference: \textit{“from the moment you see that the whole process is organized... this makes you think that the company is also like that”}. P20 added that the interviewer acts as the \textit{“company's ambassador”}, shaping the candidate's view of the entire organization.

\paragraph{CP13: Timely and Efficient Process Pacing}
An R\&S process that moves at a reasonable pace is interpreted as a sign of \textit{organizational efficiency} and \textit{respect}. P05 highlighted a particularly positive experience: \textit{“I received feedback on the same day. So, that's wonderful”}. P04 provided another concrete example: \textit{“I would do a phase on Monday. By Friday, I already had a response”}.

\paragraph{CP18: The Process as a Two-Way Street}
The R\&S process is explicitly framed as a mutual evaluation. This mindset fosters a sense of partnership, leading to attributions that the company views candidates as \textit{peers} rather than commodities. As P07 stated: \textit{“It's a two-way process, right? You are presenting yourself and the company has to present itself”}. This perspective empowers the candidate, as P10 felt: \textit{“I'm also important and I also have a voice here”}.

\paragraph{CP21: Effective Use of Automation for Efficiency}
This CP involves using automation for logistics (e.g., scheduling) without replacing human touchpoints. This balanced use is seen as a sign of a \textit{smart} and responsive organization that understands where human interaction is most valued. As P21 noted, while automation can be \textit{“very positive”}, there are parts of the process, \textit{“like... giving feedback”}, that should not be automated. 

\subsubsection{\textbf{T8. Fit Cultural}}
Assessing cultural fit is often a goal for candidates \cite{guzman2009occupational}. Our analysis shows this is not based on a single practice but is a judgment synthesized from multiple CPs. A \textit{Comprehensive Job Ad (CP01)} provides initial cues, as P10 sought a \textit{“hint in the vacancy to get an idea of the organizational climate”}. This is then further tested and validated through \textit{Direct Interaction with the Team (CP05)}, which P08 saw as essential to understand \textit{“personal aspects of the team”}. Therefore, the quality of every interaction, including \textit{Humanized Interaction (CP06)}, serves as a cultural data point that, according to P03, makes candidates \textit{“feel more comfortable”} about the work environment.

\subsubsection{\textbf{T9. Process Status}} 
We found that maintaining clear communication about a candidate’s progress is more than a procedural task, it is a defining element of a respectful hiring experience. When candidates are left uncertain about their status whether their application is under review, rejected, or simply forgotten, they interpret this silence as a signal of organizational disinterest or inefficiency. Participants consistently associated prolonged or unexplained waiting periods with anxiety, self-doubt, and disengagement. Timely updates, even when automated or delivering negative outcomes, were perceived as acts of consideration that preserved trust.

\paragraph{CP14: Continuous and Transparent Status Updates}
This CP involves proactively keeping candidates informed of their status. This pattern fosters attributions of \textit{respect} and \textit{informational justice}. 
For P04, this took the form of consistent communication after each process step, which served as the status update itself. She recalled this practice, noting that \textit{“at the end of each stage... they already gave me feedback”}. This communication is valued even when the outcome is negative, as P16 stated: \textit{“Even if... I were rejected and I had this feedback... I would say: Wow, this company here is... awesome”}.

\subsubsection{\textbf{T10. Perception of Justice}}
A perception of justice emerges as a key outcome of a well-conducted R\&S process~\cite{GILLILAND1993, PLOYHART2004}. While nearly every CP contributes to candidates’ sense of fairness, certain intentional practices function as explicit markers of organizational integrity. Participants associated fairness with processes that were predictable, criteria-driven, and consistently applied across candidates. Clear communication of evaluation criteria, constructive feedback, and timely closure were recurrently cited as indicators of procedural justice. In contrast, inconsistent treatment or unexplained decisions quickly undermined perceptions of legitimacy.

\paragraph{CP07: Visible Commitment to Diversity, Inclusion, and Accessibility}
This CP involves demonstrating a genuine commitment to Diversity, Inclusion, and Accessibility through concrete actions. These visible signals lead to attributions of an \textit{ethical} and \textit{inclusive} culture. As P05 noted, \textit{“vacancies... reserved for women... we know that women are minorities in the technology area, so that's very cool”}. This perspective extends to other groups, as P00 emphasized the fairness of allocating vacancies to \textit{“people with disabilities, black people”}, for example.

\section{Discussion}
\label{sec:discussion}
This study identified 22 CPs from interviews with novice software engineers. Interpreted through AART, these CPs offer an empirically grounded view of positive R\&S experiences. The discussion situates CPs as practical tools for designing human-centered hiring processes that better support candidate experience.


\subsection{Interpreting the CPs through the lens of AART and Related Literature}
\label{subsec:discussion_aart_cps}

Applicant Attribution-Reaction Theory (AART) helps us explain why the CPs matter, since they operate as positive ``objective R\&S events'' that shape favorable attributions of honesty, competence, and fairness toward the organization \cite{PLOYHART2004}. Their importance is clearer when viewed alongside existing literature, which has concentrated on diagnosing dysfunctions \cite{BEHROOZI2022_BADPRACTICES, MONTANDON2021} in tech hiring. CPs stand out not as optional “nice-to-haves”, but as candidate-endorsed responses to well-documented systemic problems.

A useful point of contrast is the work of \citet{SETUBAL2024}, who catalogued Anti-patterns (APs) from the recruiter’s perspective. Our study complements theirs by showing how candidate-centered CPs illuminate the other side of these dynamics. Importantly, CPs are not simply the absence or reversal of APs. While APs highlight recurring dysfunctions, CPs represent constructive practices that organizations can adopt to actively foster fairness, transparency, and engagement. CPs move the discussion from problem-diagnosis toward solution-building, offering organizations a potential  approach for designing more human-centered hiring processes. 

Recruiters lamented issues such as \textit{Inflated Experiences and Skills} and \textit{Lack of Alignment with the Job Profile} \cite{SETUBAL2024}. From the candidate’s perspective, CPs like \textit{Comprehensive and Transparent Job Advertisements} (CP01) and \textit{Relevant and Purposeful Technical Assessments} (CP04) directly address these frictions by setting clearer expectations and aligning evaluation with job demands. This overlap shows that practices perceived as fair by candidates improve recruiters’ ability to identify fit. Yet, the comparison also reveals a structural tension: recruiters prioritize efficiency and filtering, whereas candidates value developmental engagement. The emphasis participants placed on \textit{Specific and Developmental Feedback} (CP03) and \textit{Proactive Candidate Redirection} (CP15) supports this gap, highlighting that what candidates see as essential for learning and inclusion often clashes with processes optimized for speed and volume.

Furthermore, our findings provide a constructive counterpoint to the extensive body of work diagnosing flaws in technical interviews. \citet{BEHROOZI2019}, for example, detail how assessments are often stressful, detached from real practice, and perceived as unfair. While these critiques highlight what is broken, our CPs point toward how assessments can be designed differently. The combination of \textit{Relevant and Purposeful Technical Assessment} (CP4), \textit{The Interview as a Learning Experience} (CP16), and \textit{ Technically Proficient Interviewers} (CP20) outlines the elements of an evaluation that candidates interpret not as a hurdle, but as a fair, relevant, and even enriching professional exchange.

Our CPs also address challenges at the very beginning of the hiring funnel.  \citet{MONTANDON2021} showed how job ads often misrepresent actual roles, while \citet{fritzsch2023resist} documented the hype-driven dynamics of Résumé-Driven Development. Against this backdrop, participants in our study strongly valued \textit{Comprehensive and Transparent Job Advertisement} (CP01) and \textit{Co-creation of Job Descriptions with Technical Teams} (CP09), emphasizing the need for clarity and realism from the outset. These CPs represent a response to recurring mismatches between candidate expectations and organizational communication and demonstrate they are not abstract ideals but ``candidate-endorsed remedies'' to well-documented dysfunctions. Then, we may advocate that CPs could offer an actionable path for building hiring processes that balance organizational efficiency with fairness and transparency.


\subsection{Theoretical Implications for SE Research}
\label{subsec:discussion_theoretical}

This study offers theoretical contributions to the human aspects of software engineering. First, it demonstrates the value of Applicant Attribution–Reaction Theory (AART) \cite{PLOYHART2004} as an explanatory and generative framework for designing human-centered R\&S experiences in the SE context. By interpreting Constructive Patterns (CPs) as events that foster favorable attributions of fairness, competence, and respect, we show how attributional reasoning can guide the engineering of sociotechnical systems that support both organizational needs while also promoting candidate well-being.

Second, by introducing \textit{Constructive Patterns } (CPs), we add a necessary counterpart to the Anti-patterns (AP) proposed by \citet{SETUBAL2024}. Whereas APs emphasize dysfunctions in hiring, CPs provide a solution-oriented vocabulary for describing R\&S practices that candidates perceive as fair, transparent, and constructive. These two perspectives form a more balanced framework that allows SE researchers to analyze  R\&S as both a site of recurring problems and a domain where deliberate design can foster positive outcomes.

Finally, our work advocates for reframing R\&S as a critical sociotechnical system within software engineering that should be deliberately designed with human-centered principles. We refer to this perspective as \textbf{human-centered tech hiring}, in which candidate experience is treated not as optional enhancements but as core design requirement. In this regard, we advocate that candidate experience encompasses how applicants interpret and feel about the process, whether they perceive it as clear, respectful, and fair, and whether it supports rather than undermines their sense of competence and belonging \cite{carpenter2013improving, miles2018candidate}. By foregrounding human-centered tech hiring, organizations can better align hiring with long-term goals: attracting diverse talent, strengthening team cohesion, and building inclusive engineering cultures. Hiring is not merely a transactional filter but a formative sociotechnical process that actively shapes both individual careers and collective engineering capacity.

\takeawaybox{\faBookOpen \ \textit{Summary of Theoretical Implications}}{This study shows how AART can guide the design of recruitment practices that generate more positive candidate experiences in software engineering. By introducing Constructive Patterns (CPs) as a complement to Anti-patterns (APs), we contribute a vocabulary that enables both the diagnosis of problematic practices and the articulation of constructive alternatives. We argue for treating R\&S as a sociotechnical system, deserving the same analytical rigor as other SE processes, and positioning candidate experience as critical to advancing human-centered tech hiring.}

\subsection{Key Takeaways for Practitioners}
\label{subsec:discussion_practical}


Our contribution goes beyond a descriptive catalog by positioning CPs as \textit{design constructs} for human-centered tech hiring. Each CP is defined as an observable practice, grounded in empirical data, and linked to attributional mechanisms through AART, enabling systematic operationalization across R\&S stages. While this study does not empirically validate CPs through organizational deployment, it outlines pathways for use. We therefore frame CPs as empirically grounded design hypotheses to be instantiated, evaluated, and refined in future intervention-based research. We summarize the key takeaways for practitioners below.

\begin{itemize}
    \item \textbf{Build Trust Early with Principled Transparency.} Hiring begins with the job ad. Clear, honest, and technically accurate descriptions (CP01, CP09), including upfront information on responsibilities, the tech stack, and compensation, demonstrate respect, attract better-aligned candidates, and build trust from the very start.
    
    \item \textbf{Treat Every Interaction as a Cultural Signal.} Every email, update, or conversation is interpreted as a reflection of company culture (CP11). Specific and constructive feedback (CP03), timely status updates (CP14), and respectful, humanized interaction (CP06) strengthen employer reputation. Even rejections, if handled thoughtfully, can reinforce credibility.

    \item \textbf{Design Assessments to Reveal Potential, Not Just to Filter.} Technical interviews and tests should be fair and job-relevant (CP04), facilitated by competent, well-prepared interviewers (CP20), and framed as opportunities for mutual learning (CP16). These holistic evaluations (CP22) shift focus from simple gatekeeping to recognizing candidate growth and future potential.

    \item \textbf{Think Strategically Beyond the Immediate Hire.} Effective organizations treat hiring as long-term relationship-building. CPs such as framing selection as mutual evaluation (CP18) and offering constructive redirection to non-hires (CP15) contribute to stronger talent networks.

    \item \textbf{Measure, Iterate, and Adapt.} Human-centered hiring is an evolving and data-informed process. Hence, candidate satisfaction scores, direct feedback surveys, and internal audits (e.g., AP/CP self-assessments) may help organizations monitor and assess hiring performance.
\end{itemize}

\takeawaybox{\faBookOpen \ \textit{Summary of Practical Takeaways}}{
Rather than prescribing specific actions, our findings identify design principles for online human-centered tech hiring. Positive candidate experiences arise when hiring processes align expectations early, communicate intentions clearly, and frame assessments as evaluations of potential rather than exclusion. Viewing hiring as a relational process shapes perceptions of fairness and respect. These principles position candidate experience as an evolving property of the hiring system, requiring ongoing reflection rather than one-time fixes.}


\section{Threats to Validity}
\label{sec:threats}

As an exploratory qualitative study, our findings are shaped by interpretive and contextual factors. A first consideration concerns \textbf{credibility}. Thematic analysis inevitably involves researcher judgment, which means our perspectives may have influenced the definition and characterization of the CPs. In addition, participants were asked to recall past R\&S experiences, making recall bias and social desirability effects possible. We sought to mitigate these risks in different ways: grounding our AART-informed interview script in established theory, conducting regular peer debriefing sessions to challenge interpretations and build consensus, maintaining a reflexive stance throughout the analysis, and fostering a trust-based and anonymous interview environment to encourage detailed narratives. To further ground our analysis in lived experiences, the interview protocol encouraged participants to describe concrete situations rather than general opinions, and probing questions were used to delve into the specifics of their narratives. A pilot interview was also conducted to refine the script’s clarity.

In addition, the sample reflects limited demographic diversity (e.g., gender and socio-economic background) and includes participants from different professional roles (e.g., software development, QA, DevOps, UI/UX), which may affect how certain CPs are experienced or prioritized. Our focus on online tech hiring also introduces further contextual specificity that should be considered when interpreting and applying the findings in other settings.

A second consideration concerns \textbf{transferability}. Our data were collected from 22 early-career software engineers in Brazil, whose perceptions are shaped by local cultural norms, labor market conditions, and the specific characteristics of entry-level hiring. Moreover, the sample reflects limited demographic diversity (e.g., gender and socio-economic background) and includes participants from different professional roles (e.g., software development, QA, DevOps, UI/UX), which may affect how certain CPs are experienced or prioritized. Our focus on online tech hiring also introduces further contextual specificity that should be considered when interpreting and applying the findings in other settings. These factors may limit straightforward generalization to other countries, scenarios, or senior roles. Rather than claiming statistical representativeness, we aim for analytical generalization. In this regard, the core mechanisms we describe (how transparency, fairness, and respect shape candidate perceptions) are broadly relevant in global hiring contexts, as supported by related work. By providing ``thick descriptions'' and illustrative quotes, we also enable readers to judge the resonance and applicability of our findings.

Finally, we addressed \textbf{dependability and confirmability} by adopting a transparent and systematic methodology. All interviews were conducted by a single trained researcher following a interviewer consistency and theory-informed protocol \cite{braun2013successful}. Our analysis was inspired by Braun and Clarke’s \cite{braun2012thematic} six-phase thematic approach, with coding and theme development documented step by step. To support traceability and auditability, we have made available the key research artifacts (interview script, qualitative codebook, and informed consent form) in our repository \cite{zenodo}. 



\section{Concluding Remarks}
\label{sec:conclusion}

The journey into the software engineering profession is frequently shaped by flawed Recruitment and Selection (R\&S) processes. Shifting the focus from diagnosis to design, this study addressed the following research question: \textit{What constructive patterns shape positive experiences for novice software engineers in online recruitment and selection processes?} Guided by Applicant Attribution-Reaction Theory (AART), we conducted 22 interviews with early-career developers and characterized 22 Constructive Patterns (CPs) that together outline what a human-centered hiring experience can look like, from the initial job advertisement to the final offer.

In response to our research question, the findings show that CPs, ranging from comprehensive job advertisements and developmental feedback to respectful communication and mutual evaluation, are recurring practices that foster candidates experience as transparent, fair, respectful, and developmentally meaningful. In other words, they operate as positive ``objective R\&S events'', eliciting favorable attributions about the organization and reinforcing engagement and motivation. Thus, our contribution lies in introducing CPs as a solution-oriented counterpart to Anti-patterns (APs), offering both researchers and practitioners a shared vocabulary and actionable framework. Far from being optional ``nice-to-haves'', CPs emerge as strategic mechanisms that shape candidate experience and directly influence motivation to join an organization.

For the industry, the findings translate into a call to treat the candidate experience with the same human-centered rigor and creativity applied to software engineering. This socially constructed process involves, for instance, building trust through clear and transparent job ads, treating every communication as a cultural signal, designing assessments that reveal potential rather than simply filtering, and approaching hiring as the beginning of a long-term professional relationship. These steps may be advantageous in attracting and retaining talent in a competitive labor market. 

This study points to future research on the feasibility, cost, and effectiveness of CPs in real hiring pipelines; their identification, prevalence, and impact across diverse samples; and the causal effects of interventions on candidate and organizational outcomes. Comparative and longitudinal work could examine cultural, contextual, and seniority differences, including how early hiring experiences shape careers. As hiring becomes more algorithmic, research should also assess how CPs interact with AI, automation, and bias in R\&S.

\section*{ARTIFACTS AVAILABILITY}
To promote transparency and reproducibility, all materials supporting this study are available in our public repository \cite{zenodo}.


\begin{acks}
We thank the 22 participants and the reviewers for their contributions. This study was supported by Brazilian Federal Agency for Support and Evaluation of Graduate Education - CAPES (Postdoctoral Institutional Program – PIPD), Brazilian Research Council - CNPq (Universal Grant 404406/2023-8; Grant 312275/2023-4), Rio de Janeiro State's Research Agency - FAPERJ (Grant E-26/204.256/2024), the Kunumi Institute, Stone Co, and TNE-DeSK (CUP H91I24000380007). We also acknowledge institutional support from the UFCA, Saarland University, the University of Bari, and the PUC-Rio.

\end{acks}

\bibliographystyle{ACM-Reference-Format}
\bibliography{references}

\end{document}